June 23$^{nd}$, 2011

# Possible, alternative explanations of the T2K observation of the $\nu_e$ appearance from an initial $\nu_\mu$.

**D. Gibin**[(a)], **A. Guglielmi**[(a)], **F. Pietropaolo**[(a)], **C. Rubbia**[(b)(c)] and **P. Sala**[(d)]

[(a)]*Dipartimento di Fisica e INFN, Università di Padova, Padova, Italy*
[(b)]*Laboratori Nazionali del Gran Sasso dell'INFN, Assergi (AQ), Italy*
[(c)]*CERN, Geneva Switzerland*
[(d)]*INFN, Sezione di Milano, Milano, Italy*

ABSTRACT: An alternative explanation to the emergence of $\sin^2(2\theta_{13}) > 0$ is discussed. It is pointed out that the recorded T2K events might have been due to some other new physics in the neutrino sector, related to the LSND/MiniBooNE sterile neutrino anomalies, for which there is nowadays a growing evidence. The presently running ICARUS detector with the CNGS beam will be able to distinguish between these two possible sources of the effect.

The T2K collaboration has just reported [1] the indication of the $\nu_e$ appearance after 295 km from initial $\nu_\mu$. Based on the 2.5° off-axis orientation of the neutrino beam and the known value of $\left|\Delta m_{23}^2\right| = 2.4 \times 10^{-3} eV^2$, $\sin^2(2\theta_{23}) = 1$ the beam energy for the oscillation first maximum is $E_\nu = 600 MeV$, near the chosen optimum neutrino energy. A total of 6 $\nu_e$ candidates have been recorded out of a total of 121 fully contained neutrino events. The expectations for the standard three-flavour neutrino oscillation scenario with $\sin^2(2\theta_{13}) = 0$ are of 0.8 events for the beam associated $\nu_e$ background and of 0.6 events from the neutral current induced background. Added to the 0.1 events due to $\nu_\mu \to \nu_e$ oscillated solar term, the prediction is then of $1.5 \pm 0.3$ events. The probability to observe these 6 or more events with the expectation of 1.5 events is $7 \times 10^{-3}$, equivalent to a 2.5 σ significance. The T2K collaboration has analysed the data within the three-flavour neutrinos assumption, $\sin^2(2\theta_{13}) > 0$ and $\delta_{CP} \approx 0$ with the result $0.03(0.04) < \sin^2(2\theta_{13}) < 0.28(0.34)$ for normal (inverted) hierarchy at 90% confidence level. A previous search by CHOOZ [2] had already excluded the $\bar{\nu}_e$ disappearance for the standard neutrino scenario with a limit $\sin^2(2\theta_{13}) > 0.14$ at 90% confidence level.

We would like hereby point out that the alternative $\sin^2(2\theta_{13}) > 0$ may be by no mean the only possible origin of the above observed excess of oscillated $\nu_\mu \to \nu_e$ events [1]. In particular we are investigating if the above mentioned (6 − 1.5) = 4.5 $\nu_e$ events might have been related to some other new physics in the neutrino sector, for which there is nowadays some growing evidence.

As well known, a first observation of an anomalous oscillated $\bar{\nu}_\mu \to \bar{\nu}_e$ excess with much larger values of mass difference $0.4 < (L/E_\nu) < 1.2$ $(meters/MeV)$ had been reported by the LNSD experiment [3] at LANSCE and a strong 800 MeV, 1 mA proton beam, with the signature of Cerenkov light from $e^+$ from positive pion and muons decays at rest and a delayed neutron-capture [n+p $\to$ d+$\gamma$ (2.2 MeV)]. Since the intrinsic beam related $\bar{\nu}_e$ rate is only 4 x $10^{-4}$ relative to $\bar{\nu}_\mu$, a significant $\bar{\nu}_e$ rate would be evidence for $\bar{\nu}_\mu \to \bar{\nu}_e$ oscillations. An excess of (87.9 ± 22.4 ± 6.0) events with a strong, 3.8 σ significance has been reported. The LNSD result, by now some fifteen years old, is still unchallenged.

The subsequent experiment MiniBooNe [4] is looking at the excess of (anti-) $\nu_e$ events in a (anti-)$\nu_\mu$ beam from the 8 GeV proton Accelerator at FNAL. Significant $\nu_e$ (~3 σ) and $\bar{\nu}_e$ (~2.5 σ) excesses above background are emerging in both neutrino mode and antineutrino mode. While the excess for the process $\bar{\nu}_\mu \to \bar{\nu}_e$ is in good agreement with the LNSD result, for $\nu_\mu \to \nu_e$ the excess has a different $(L/E_\nu)$ distribution, indicating perhaps the existence of a more complex situation. A number of alternatives have been postulated [5]. For instance in order to explain such a difference for instance within the sterile neutrino models, two additional sterile neutrinos are described with independent $\Delta m^2$, and mixing parameters and with differences between neutrino and antineutrino due to the presence of a new CP-violating phase. The $\nu_e$ appearance from initial $\nu_\mu$ would then be described as:

$$P(\overset{(-)}{\nu_\mu} \to \overset{(-)}{\nu_e}) = 4|U_{\mu 4}|^2 |U_{e4}|^2 \sin^2(\theta_{41}) + 4|U_{\mu 5}|^2 |U_{e5}|^2 \sin^2(\theta_{51}) + $$
$$8|U_{\mu 4}||U_{e4}||U_{\mu 5}||U_{e5}|\sin(\theta_{41})\sin(\theta_{51})\cos(\theta_{54} \pm \varphi_{45})$$

where $\varphi_{45}$ is an additional CP violating phase, with opposite signs respectively for the neutrino and antineutrino alternatives. However the presence of a considerable "tension" in the phenomenological analysis of the whole set of neutrino oscillation experiments must be acknowledged [5] [9].

Recently significant additional anomalies in the neutrino disappearance rates have been observed. Too few neutrino interactions are observed when compared to the values predicted from the source. They are (a) the Reactor anomaly [6] in the $\bar{\nu}_e$ data, where $\nu_{measured}/\nu_{expected} = 0.943 \pm 0.023$; (b) the Gallium source anomaly [7] in the $\nu_e$ data, where $\nu_{measured}/\nu_{expected} = 0.86 \pm 0.05$. These new results have further increased the interest in the possibility that additional sterile neutrinos might exist.

The LSND [3] and the subsequent MiniBooNe [4] experiments of electron neutrino appearance from an initial muon neutrino beam are usually described within the framework of a two neutrino mixing with $P_x = \sin^2(2\theta_x)\sin^2(1.27\Delta m_x L/E_\nu)$. The $\nu_e$ appearance from initial $\nu_\mu$ reported by T2K [1] could then be instead associated to the above described oscillatory behaviour, strongly averaged because of the very long flight path distance, $P_x(\nu_\mu \to \nu_e, L \gg E_\nu/\Delta m_X) \approx 0.5\sin^2(2\theta_x)$. The straightforward analysis of data of T2K [1] with $\sin^2(2\theta_{13}) = 0$ will then correspond to $0.06 < \sin^2(2\theta_x) < 0.54$ at 90% confidence level.

In order to provide a first evidence for the range of estimated mass differences, the T2K result may be compared to the LNSD result — however for anti-nu — $P(\bar{v}_\mu \to \bar{v}_e) = (0.245 \pm 0.067 \pm 0.045)\% = \sin^2(2\theta_x)\sin^2(1.27\Delta m_x L/E_v)$, as shown in Figure 1 where the T2K result is superimposed to the LSND/MiniBooNE allowed regions in the in the $\Delta m_x^2 - \sin^2(2\theta_x)$ parameter space.

The experiment CNGS2 presently running at the LNGS with high-energy neutrinos coming from CERN may be able to distinguish between the two above described options of T2K. This detector, named ICARUS [8], consists of about 600 tons of high purity LAr in an "electronic bubble chamber" and it is capable of searching after 732 km with a very high degree of confidence $v_e$ coming from the 18 GeV average energy wide band $v_\mu$ beam initially produced by the 400 GeV CERN-SPS.

Both LSND and MiniBooNE liquid scintillator experiments are characterised by quasi-elastic events $\leq \approx 1$ GeV on Carbon with elaborate selection criteria, a relatively short oscillation path from the source to the detector and with substantial competing backgrounds due to poor $e - \pi_o$ separation, in which the "signal" appears as an excess of events.

In contrast, an equivalent search with the ICARUS detector with CNGS2 is based on deep inelastic $v_e$ CC events recorded in the minimum bias mode with high efficiency and an extremely good background discrimination (NC rejection > $10^3$), limited only by the intrinsic beam $v_e$ contamination $\leq 0.5\%$ in the 10 ÷ 30 GeV neutrino window. Like T2K, CNGS2 operates on an oscillation path much longer than the ones of LNSD/MiniBooNE, observing an averaged signal for a LNSD anomaly. The very long path-length ensures several oscillations from source to detector in the foreseen LSND window, but still wide enough in order to identify maxima and minima related to LSND-like signal, given the high accuracy of the energy resolution of the (contained) events. Therefore in addition the actual value of such a new mass difference may also be investigated.

Notwithstanding, in contrast with T2K, the expectations for the standard three-flavour neutrino oscillation scenario with $|\Delta m_{23}^2| = 2.4 \times 10^{-3} eV^2$, $\sin^2(2\theta_{23}) = 1$ give a negligible contribution to the expectations for the alternative $\sin^2(2\theta_{13}) > 0$ and a very small contribution to $v_\mu \to v_\tau$ production with an off-axis electron event and we may distinguish between the alternatives.

In particular if the net effect of the (6 -1.5) = 4.5 events of T2K experiment were <u>entirely</u> due to LNSD-like $v_\mu \to v_e$ events, with the ≈1500 fully contained neutrino events expected by the end of 2011 run we would record of the order of 4.5/121 x 1500 ≈ 56 oscillated $v_e$ events, to be compared with a intrinsic beam related background of ≈ 7.5 events. The ICARUS search is instead insensitive to the $\sin^2(2\theta_{13})$ option of the T2K result because of the much higher energies of the CNGS neutrino beam.

We gratefully acknowledge the ICARUS collaboration [8] for critical reading of the manuscript and for their extremely useful suggestions.We gratefully acknowledge the ICARUS collaboration [8] for critical reading of the manuscript and for their extremely useful suggestions.

# References (rev.)

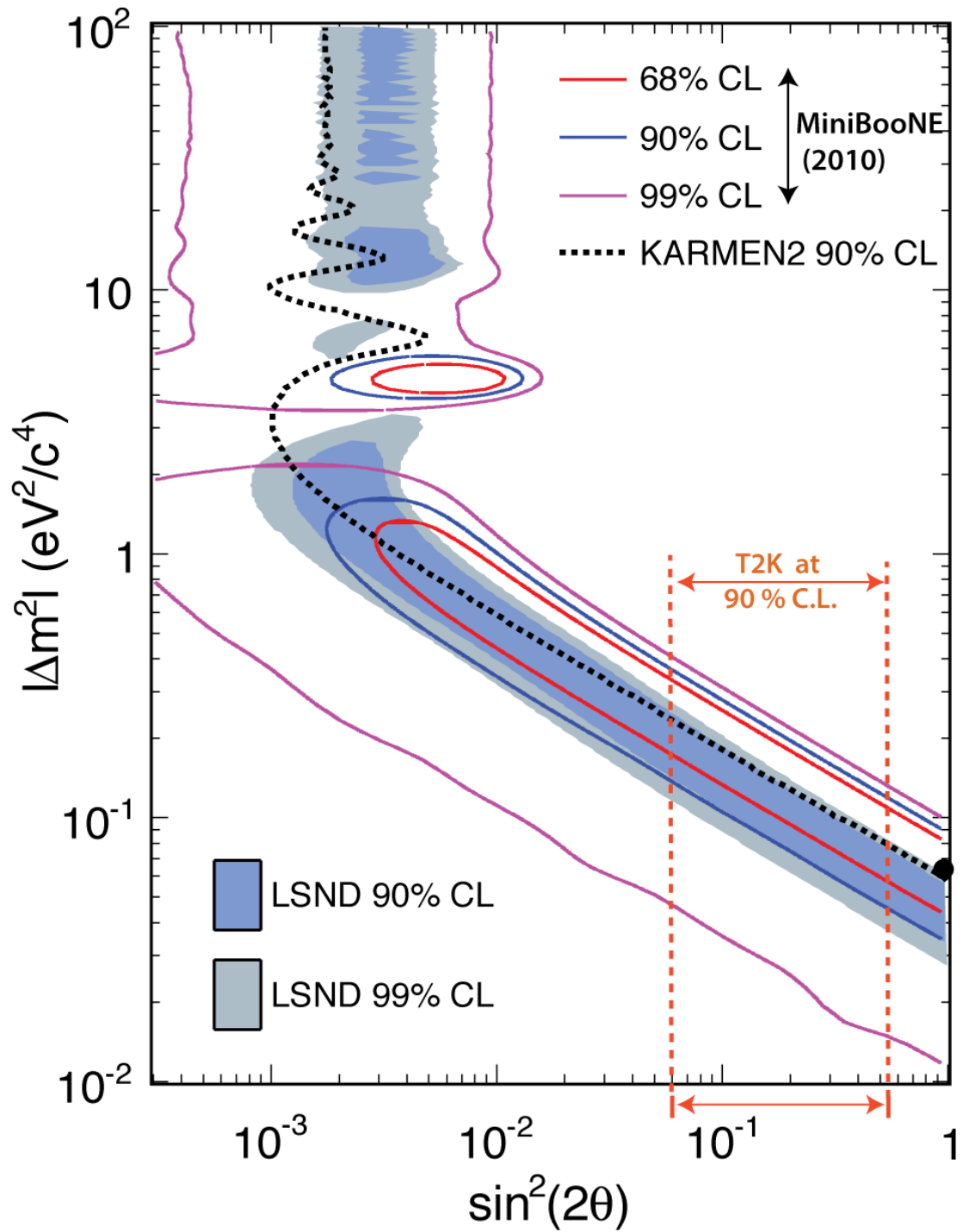

**Figure 1.** Allowed regions in the [$\sin^2(2\theta_x) - \Delta m_x^2$] plane for the LNSD experiment at 90% and 99% confidence levels and for the MiniBooNE experiment for incoming anti-neutrino (from Reference [4]). The prediction for the T2K experiment at 90% confidence level is also shown. It must be remarked that the last prediction may not be immediately applicable to the previous ones, since the T2K experiment is based on incoming neutrinos.